# RF CELL MODELING AND EXPERIMENTS FOR WAKEFIELD MINIMIZATION IN DARHT-II

Scott D. Nelson[†] and Michael Vella[‡]

Lawrence Livermore National Laboratory, Livermore, California 94550 USA[†]

Lawrence Berkeley National Laboratory, Berkeley California 94720 USA[‡]

## *Abstract*

Electron beams of linear induction accelerators experience deflective forces caused by RF fields building up as a result of accelerating cavities of finite size. These forces can significantly effect the beam when a long linac composed of identical cells is assembled. Recent techniques in computational modeling, simulation, and experiments for 20 MeV DARHT-II (Dual Axis Radiographic Hydrodynamic Test) accelerator cells were found to reduce the wakefield impedance of the cells from 800 ohms/meter to 350 ohms/meter and experimental results confirm the results of the modeling efforts. Increased performance of the cell was obtained through a parametric study of the accelerator structure, materials, material tuning, and geometry. As a result of this effort, it was found that thickness-tuned ferrite produced a 50% deduction in the wakefield impedance in the low frequency band and was easily tunable based on the material thickness. It was also found that shaped metal sections allow for high-Q resonances to be de-tuned, thus decreasing the amplitude of the resonance and increasing the cell's performance. For the geometries used for this cell, a roughly 45 degree angle had the best performance in affecting the wakefield modes.

## 1 INTRODUCTION

The modeling, simulation, design, and experimental activities for the DARHT-II accelerator cell (Figure 1) consisted of several stages in order to lower the wakefield impedance of the cell while maintaining the voltage breakdown hold off characteristics of the pulsed power design. For the computational simulation and RF design of the cell, the AMOS[1] code was used.

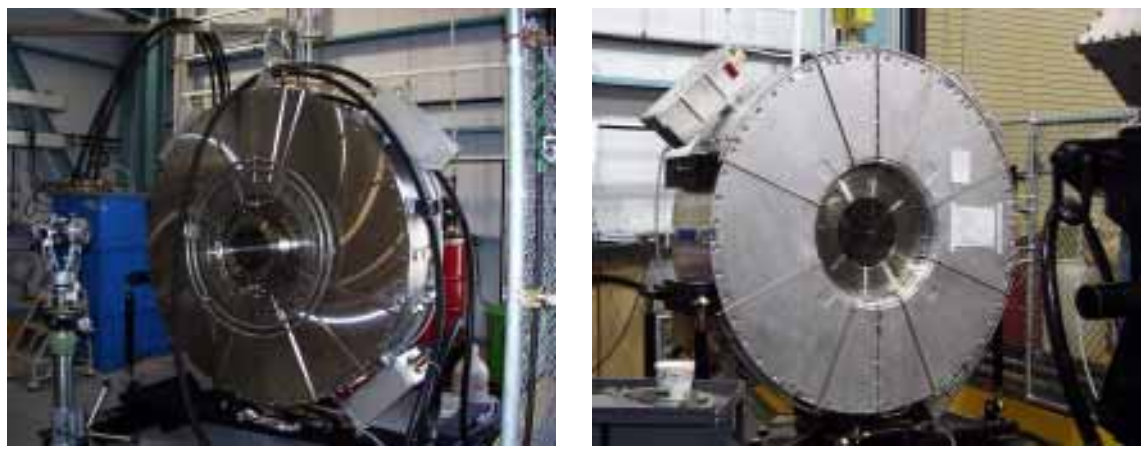

*Figure 1. The DARHT-II [standard] accelerator cell (left) and the DARHT-II injector accelerator cell (right) are shown during high voltage and RF (radio frequency) tests at LBL.*

The inside of the accelerator cell consisted of an accelerating gap, a Mycalex™ insulator, a metglas core, and was oil filled. As part of the design and simulation activities for the cell, the material properties of the ferrite and Mycalex used in the cell were obtained and a quick comparison between the simulation code and an ideal *pillbox* cavity was generated to study the gridding effects in the results. This was especially important since the ferrite tuning effort was very sensitive to thickness variations. The accelerator cell was fabricated to match the resulting simulation results and both [standard] accelerator cells and injector accelerator cells were tested experimentally[2].

## 2 SIMULATION OF THE CELL

The AMOS 2.5D FDTD (finite difference time domain) RF code was used to simulate the cross section of the cell.

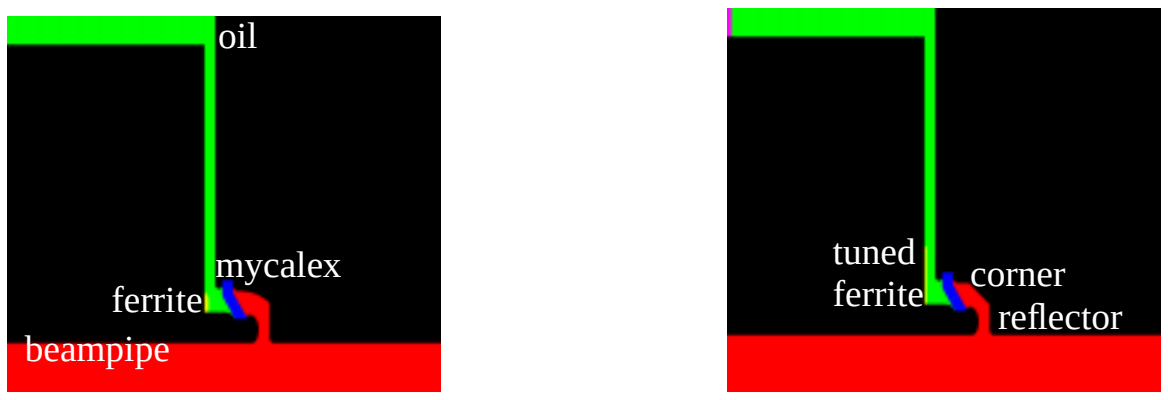

*Figure 2. The AMOS model of the original DARHT-II standard accelerator cell (left) includes a Mycalex insulator $\varepsilon_r$=6.8 to restrict the oil. The final design modifications (right) contain similar structures as (left) but with the addition of the thickness tuned ferrite in the oil region and the flat corner reflector in the vacuum region.*

Initial RF experimental results from the cell had poor performance and the first models had peaks at 800Ω/m so design efforts commenced to address these issues (see Figure 6.) The resonances were caused by standing waves in the cell in high-Q regions of the cell. These high-Q regions in the model were then identified via temporal-spatial Fourier transforms. This identified several properties of the original cell: the high-Q regions in the cell are in the area of the accelerating gap and in the vertical oil column; very little energy traverses the upper corner of the structure; and the performance is very sensitive to the gap geometry. Due to fabrication issues, it was not possible to consider changes to the Mycalex piece. So subsequent design changes to the cell concentrated only on modifications to metal pieces or the addition of metal inserts to the cell. This also allowed for easy cell modification.

As part of the simulation activities it was desirable to reduce the high-Q resonance that occurs at the low frequency point near 200 MHz. The tuning of the thickness of the ferrite tiles in the oil region were found to be a very effective way to reduce this resonance. But simulations of these tiles are very sensitive to the tile thickness since it was instrumental in the tuning process. Since the performance of the simulation code with these tuned ferrites was critical, a study looked at the errors based on cell size.

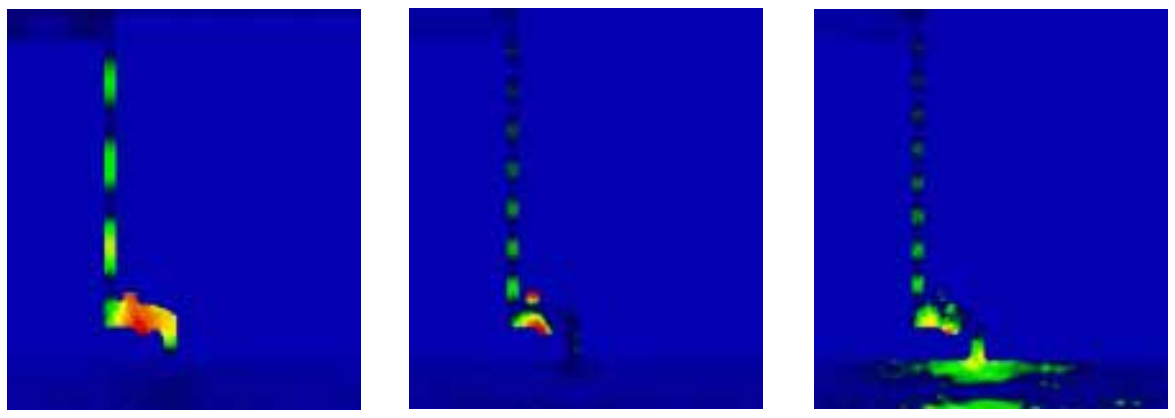

*Figure 3. The temporal-spatial Fourier transform shows the locations of the high-Q resonances (refer to Figure 1 for the geometric layout of the components). The image on the left is the low frequency resonance (see Figure 6), the image in the middle is the high frequency resonance, and the image on the right is the pipe mode (for confirmation purposes).*

The *pillbox* equation from [3] with the appropriate scale factors to correspond to the AMOS calibrations in MKS units has a $8.987\times10^{11}$ scale factor to convert into Ω/m. A source of error was found to be the meshing process which changed the geometry of the solved problem — i.e. as a result of the meshing, the geometry that was solved was different than the initial geometry. Since the ferrite tuning effort was very sensitive to ferrite thickness, it was important to select cell sizes such that the ferrite and the gap contain an integer number of cells. Failure to do so causes an almost 20X increase in the error of the calculation (see Figure 4).

$$Z_\perp(\omega) = -\frac{2cdZ_0}{\pi\omega b^2}\Im\left\{\frac{\left(\sin\frac{\omega d}{v}\right)^2}{\left(\frac{\omega d}{v}\right)^2}\frac{1}{H_1(\omega)}\right\}\times 8.987\cdot 10^{11}$$

$$H_1(\omega) = \frac{\omega b}{c}\left(\frac{J_1'(\omega b/c)}{J_1(\omega b/c)} - \frac{G_1'(\omega, b)}{G_1(\omega, b)}\right) - \frac{1}{d}\sum\frac{1-e^{-2d\mu_s(\omega)}}{\mu_s(\omega)(\rho_s{}^2 b^2 - 1)} + \frac{1}{d}\left(\frac{\omega}{c}\right)^2\sum\frac{1-e^{-2d\varepsilon_s(\omega)}}{\varepsilon_s(\omega)^3}$$

$$Y_1'(\omega b/c) = \partial_x Y_1(x)\big|_{\omega b/c},\quad J_1'(\omega b/c) = \partial_x J_1(x)\big|_{\omega b/c}$$
$$G_1(\omega, b) = J_1(\omega b/c) + C_1(\omega)\times Y_1(\omega b/c)$$
$$G_1'(\omega, b) = J_1'(\omega b/c) + C_1(\omega)\times Y_1'(\omega b/c)$$

$$\mu_s(\omega) = \sqrt{\rho_s^2 - (\omega/c)^2},\ \varepsilon_s(\omega) = \sqrt{\beta_s^2 - (\omega/c)^2}$$

$$C_1(\omega) = -\frac{j(Z_s/Z_0)J_1'(\omega R/c) - J_1(\omega R/c)}{j(Z_s/Z_0)N_1'(\omega R/c) - N_1(\omega R/c)}$$

where $\rho_s, \beta_s$ are 1/b × the zeros of $J_1', J_1$, $J_1$ and $Y_1$ are the Bessel functions of the first and second kind, $\Im\{x\}$ is the imaginary operator, $c$ is the speed of light, $Z_0=4\pi/c$, $Z_s=2*Z_0$, $v=\beta c$, and $b$, $d$, $R$ are the beampipe radius, the *pillbox* longitudinal half-width, and the *pillbox* radius respectively. The error with-out regard to the geometry was 6%. Taking into account the geometry resulting from the meshing process reduced the error to 0.35%.

## 3 MATERIAL TESTING

As part of the construction of the simulation model, the properties of the Mycalex and ferrite were measured using the coaxial line.[4,5,6] This involves acquiring the [s] parameters for the sample in a coaxial line and then applying the following relationships:

$$V_1 = S_{21} + S_{11},\ V_2 = S_{21} - S_{11},$$

$$X = \frac{1 - V_1V_2}{V_1 - V_2},\ \Gamma = X \pm \sqrt{X^2 - 1},\ Z = \frac{V_1 - \Gamma}{1 - V_1\Gamma}$$

the sign choice for $\Gamma$ is resolved by requiring $|\Gamma| \le 1$. Relative permeability and permittivity are then obtained as $\mu_r = \sqrt{c_2c_1},\ \varepsilon_r = \sqrt{c_2/c_1}$, where

$$c_1 = \frac{\mu_r}{\varepsilon_r} = \left(\frac{1+\Gamma}{1-\Gamma}\right)^2$$

$$c_2 = \mu_r\varepsilon_r = -\left[\frac{c}{\omega d}\ln\frac{1}{Z}\right]^2$$

Using these relations, the Mycalex was measured to have a permittivity of $\varepsilon_r = (6.8 - j0.5) \pm (0.2 + j0.2)$ and a permeability of $\mu_r = (1.0 + j0.5) \pm (0.1 + j0.2)$ up to 800 MHz. The error bars were determined from a sample of teflon of the same size; but it should also be noted that the technique is dependent on standing wave patterns formed in the line, thus different materials have different errors. Further refinement of the technique is required.

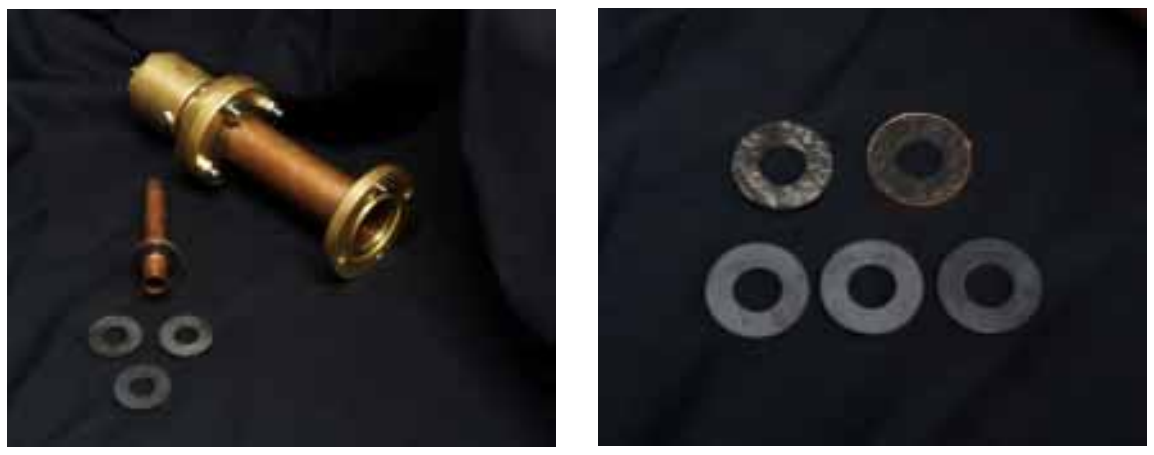

*Figure 5. The coaxial air-line test stand used for testing the material samples is shown on the left. The pipe is a 50Ω line, has type-N connectors on each end, and has a removable center conductor. The torque of each bolt was set to 50 oz.-in. and was required for calibration purposes.*

To be useful to the AMOS code, the permeability has to be expanded using a Lorentzian model:

$$\mu_r(f) = 1 + \chi_m(f)$$

$$= 1 + \frac{1}{2}\sum_{l=1}^{m}\frac{\alpha_l}{(\beta_l - \gamma_l) - j2\pi f} - \frac{\alpha_l}{(\beta_l + \gamma_l) - j2\pi f}$$

yielding these coefficients, valid for 1MHz - 1 GHz: $\alpha_1$= $2.178\times10^{10}$, $\beta_1$= $1.275\times10^{7}$, $\gamma_1$= $3.390\times10^{11}$, $\alpha_2$= $2.571\times10^{10}$, $\beta_2$= $7.375\times10^{7}$, $\gamma_2$= $6.190\times10^{10}$, $\alpha_3$= $5.574\times10^{10}$, $\beta_3$= $2.437\times10^{8}$, $\gamma_3$= $1.880\times10^{10}$.

## 4 RESULTS

The redesign activities of the cell (see Figure 2) improved the wakefield performance of the cell by 2X of the dipole mode by lowering the Q of the structures. The results for the injector cell further benefited from the increase beampipe bore size (14" vs. 10") thus dramatically reducing the high frequency resonance since it is above cutoff in a 14" bore. The monopole mode for this structure is extremely small and the quadrupole mode ($6000\Omega/m^2$) does not steer the beam and so is not detrimental to the cell's performance.

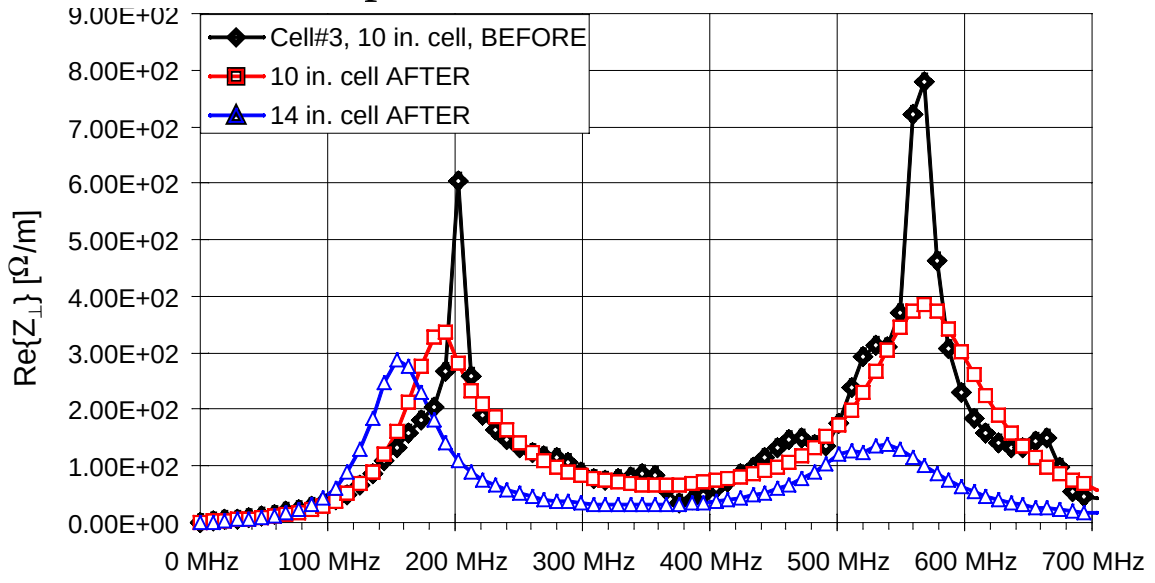

*Figure 6. The dipole-mode simulation of the 10" cell before modification is shown with black diamonds, the results after modification are shown with red squares, and the results for the 14" injector cell after modification are shown with blue triangles.*

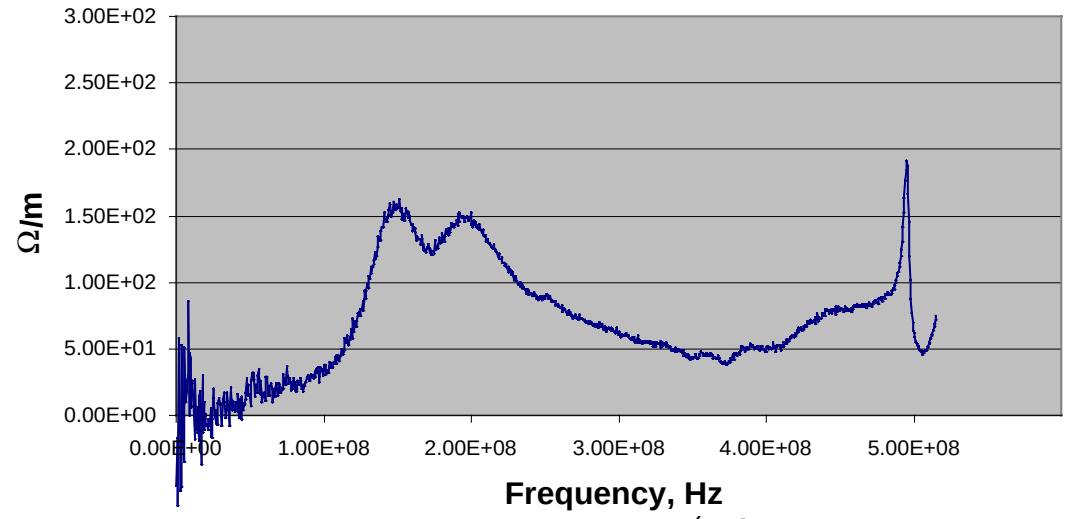

*Figure 7. The experimental data[7] from the injector cell has a lower amplitude than expected caused by a split peak at the low frequency point. The high-Q mode at ~500MHz is the pipe resonance.*

## 5 CONCLUSIONS

The redesign of the accelerator cell gap region improved the characteristics from the original 800Ω/m down to 350Ω/m. As part of this analysis, the following relationships were observed: tuning the ferrite thickness of the ferrite in the oil region reduces $Re\{Z_\perp\}$ by 60Ω/m / mm (low freq.) and 140Ω/m / mm (high freq.). Ferrite 5-6 mm thick was selected and to be fully tuned, the ferrite was placed at normal incidence to the wavefronts. A corner reflector in the vacuum region near the gap affects performance by 50-100Ω/m per 30°. The 45° point is "close" to the optimum angle in this design. Ferrite extending up the oil gap reduces resonances by 20Ω/m / cm (low freq.) as long as the fields remain normal to the surface. 13-15 cm was chosen as the best compromise between performance and assembly efforts. Changing the insulator angle reduces resonances by 100Ω/m / 10° (high freq.) but the insulator also has high voltage constraints and so the authors did not change the insulator configuration.

From the comparisons with the analytic *pillbox* models, it was determined that most of the error in the AMOS calculations is caused by gridding effects when the discretization of the mesh doesn't match the physical geometry. For those cases where the *pillbox* geometry was assigned to match the cell-generated geometry, the error was less than 0.35%. These effects are especially pronounced due to the sensitivity of the performance of the cell vs. the tuned ferrite thickness.

From the modeling and experiments on the injector cell, it was observed that the injector cell resonances scale with pipe radius. Also, the high frequency resonance seen with the standard cell is virtually eliminated with the injector cell due to the larger beampipe bore size for the injector cell thus supporting a pipe mode at the high frequency point.

## 6 ACKNOWLEDGMENTS

Thanks go to Dick Briggs for his comments on the experimental results, to Bill Fawley for his insights in the modeling efforts, to David Mayhall for his work on the *pillbox* verification, to Jerry Burke for his material parameter extraction, to Jim Dunlap for his experimental measurements, and to Brian Poole for his input on temporal-spatial Fourier transform techniques for identifying resonances. The authors would also like to thank, acknowledge, and morn the passing of Dan Birx who substantially contributed to the experimental effort and reviewed the modeling and simulation efforts. This work was performed under the auspices of the U.S. Department of Energy by the LLNL under contract No. W-7405-Eng-48.

## 7 REFERENCES

[1] J. DeFord, S. D. Nelson, "AMOS User's Manual," *Lawrence Livermore National Laboratory*, April 13, 1997, LLNL UCRL-MA-127038.

[2] S. D. Nelson, J. Dunlap, "Two-wire Wakefield Measurements of the DARHT Accelerator Cell," *Lawrence Livermore National Laboratory*, 1999, LLNL UCRL-ID-134164.

[3] R. J. Briggs, D. L. Birx, G. J. Caporaso, V. K. Neil, and T. C. Genoni, "Theoretical and Experimental Investigation of the Interaction Impedances and Q Values of the Accelerating Cells in the Advanced Test Accelerator," *Particle Accelerators*, Vol. 18, 1985, pp 41-62.

[4] G. Burke, "Broad-band Measurements of Electrical Parameters for Ferrite and Dielectric Samples," *Lawrence Livermore National Laboratory*, 2000, as yet unpublished.

[5] J. Baker-Jarvis, M. D. Janezic, J. H. Grosvenor, Jr., and R. G. Geyer, "Transmission/Reflection and Short-Circuit Line Methods for Measuring Permittivity and Permeability," *U. S. Dept. of Commerce*, NIST Technical Note 1355-R, Dec. 1993.

[6] A. M. Nicolson and G. F. Ross, "Measurement of the Intrinsic Properties of materials by Time-Domain Techniques," *IEEE Trans. Instrumentation and Measurement*, Vol. IM-19, No. 4, Nov. 1970.

[7] R. J. Briggs, M. Vella, S. D. Nelson, to be published report on the magnetic field wakefield measurements of the DARHT-II injector cell.